\documentclass[aps,prl,twocolumn,groupedaddress]{revtex4}

\usepackage{graphicx}
\newcommand{\be}{\begin{equation}}
\newcommand{\ee}{\end{equation}}

\begin{document}
\title{Ground state and thermal properties of a lattice gas 
on a cylindrical surface}
\author{M. Mercedes Calbi$^{a}$, Silvina M. Gatica$^{a,c}$, Mary J. Bojan$^{b}$, and Milton W. Cole$^{a}$}
\affiliation{Departments of $^{a}$Physics and $^{b}$Chemistry, Pennsylvania State University, University Park, Pennsylvania 16802\\
$^{c}$Departamento de F\'{\i}sica, Universidad de Buenos Aires, Buenos Aires 1428, and CONICET, Argentina}

\date{\today }

\begin{abstract}

Adsorbed gases within, or outside of, carbon nanotubes may be analyzed
 with an approximate model of  adsorption on lattice sites situated on
 a cylindrical surface. Using this model, the ground state energies of
 alternative lattice structures are calculated, assuming Lennard-Jones
 pair interactions between the particles. The resulting  energy and
 equilibrium structure are nonanalytic functions of radius (R) because
 of commensuration  effects associated with  the cylindrical geometry.
 Specifically,  as  R  varies,  structural transitions  occur  between
 configurations differing in the  "ring number", defined as the number
 of atoms  located at  a common value  of the  longitudinal coordinate
 (z). The  thermodynamic 
 behavior of  this  system  is evaluated  at
 finite  temperatures,  using   a  Hamiltonian  with  nearest-neighbor
 interactions.   The  resulting  specific  heat  bears  a  qualitative
 resemblance to that of the one-dimensional Ising model.

\end{abstract}

\maketitle

\section{Introduction}

Atoms or  molecules may be confined  either within, or  on the outside
surface  of,  cylindrical materials,  such  as  carbon nanotubes.  The
existence  of  such systems  raises  a  set  of interesting  questions
concerning  the  thermal, structural  and  dynamical  properties of  a
cylindrical monolayer film. In this paper, we evaluate the thermal and
structural properties of such a  film, with the help of two simplified
models. We  believe that some  of our results are  realistic, although
others  may be  artifacts of  the model \cite{rivista,axial}. For 
example,  Figure 1, taken  from previous  work \cite{axial}, depicts 
the  density of  H$_2$ molecules within a  nanotube of radius R=7 \AA$,$ 
at temperature T=10 K.  Note that the
radial spread of  the so-called ``cylindrical shell'' phase is some 
10\% of the  mean radial distance $\langle r \rangle \approx 3.8$ 
\AA. In such a 
situation, the  model of confinement to a precise value of R would seem 
appropriate.

The first  task we undertake is  to ascertain the  ground state energy
 and structure of an ensemble  of atoms, assumed to be classical, that
 interact  with  all  other  atoms  with a  Lennard-Jones  (L-J)  pair
 potential: $U(r)= 4\epsilon[(\sigma/r)^{12}-(\sigma/r)^6]$. Here 
$\sigma$ is the nominal 
diameter
 of the atoms and $\epsilon$ is  the well-depth of their mutual 
interaction. We
 assume, without  proof, that the equilibrium structure  is a periodic
 crystal. To  determine its properties,  we minimize the  ground state
 energy per particle in this cylindrical surface lattice. In so doing,
 we consider possible  periodic structures and determine that structure 
which has
 the lowest energy, at any specified value of R. The second problem is 
to evaluate the
 thermal properties of such a ``lattice gas'' which has varying fractional site occupancy $\theta$. In this case, we simplify
 the problem by including just nearest neighbor interactions.

The behavior  of the  system can  be expressed in  terms of  a reduced
radius  $R^*= R/\sigma$. Both the  ground  state and  finite temperature  
(T) problems  have  much-studied   one-dimensional  (1D)  and  2D  limits,
corresponding to R* = 0 and $\infty$, respectively. Interestingly, 
the behavior does  not   interpolate  smoothly  between  these  limits 
as  R*  is
varied. This  happens because \cite{green} of a commensuration  effect 
arising
from  the periodicity  associated  with the  azimuthal  angle $\phi$.  This
phenomenon is analogous to that found in monolayer films in the regime
where  the  film's  lattice  constant   is  similar  to  that  of  the
substrate \cite{millot}. In  the  cylindrical case,  the  circumference 
of  the
cylinder provides  the length scale that  determines the compatibility
of candidate structures. 

In the  next section, we evaluate the ground
state  problem by  considering  a rather  general  set of  alternative
structures. In  Section III,  we compute the  specific heat  for several
values  of the  ring number.  Finally, in  Section IV  we  estimate the
quantum effects  (by including zero-point energy) present  in a rather
extreme  case, $^4$He  atoms on  a cylindrical  surface, and  compare the
resulting energy with the known energy.

\section{Ground state energy}

We  assume  that  the  ground  state  structure  is  close-packed,  as
exemplified in Figure 2, and that it can be derived with the following
algorithm.  At any  specific value,  say zero,  of the  coordinate (z)
parallel  to  the cylinder's  axis,  there  are $\nu$ atoms,  distributed
uniformly in  azimuthal angle $\phi$. We  call $\nu$ the  ring number 
of the
structure and consider  structures with integral value of $\nu$. 
Figure 2
depicts the  case $\nu=4$, which turns  out to be an  important example. A
unit  cell  of  this structure  consists  of  four  atoms at  z=0  (at
azimuthal  angles  $\phi$=0, $\pi$/2, $\pi$ and  $3\pi/2$) and four atoms 
at  z=a ($\phi=\pi/4$, $3\pi/4$, $5\pi/4$ and $7\pi/4$). The  structure is  characterized  by a
one-dimensional density $\rho=4/a$, since there  are 8 atoms in a unit cell
of length $2a$.    For
each   such   hypothetical    structure,   we   have   performed   two
calculations.  The  first is  a  total  energy  calculation, aimed  at
determining the  lowest energy structure.  The energy in this  case is
taken as  the sum  of two-body L-J  interactions between atoms  at all
lattice sites.  The second study yields  the thermodynamic properties,
described in the next section of this paper.

Before embarking on these calculations, we assess the models. The only
approximations  in  the  ground  state  calculation  are  the  use  of
Lennard-Jones  potentials,  the omission  of  kinetic  energy and  the
assumption   that  the   actual  structure   fits   the  close-packing
description.  The first  two  are conventional  approximations in  the
lattice gas approach;  we note only that many-body  corrections may be
important  in this  geometry (but  we ignore  them)  \cite{milen}. The
third assumption seems logical,  since all simple close-packed lattice
structures are included. One  can imagine other possibilities, such as
one in which the unit cell  consists of more than two rings, but these
seem  implausible to  us. We  note that  the problem  of packing  on a
spherical surface is  quite different from that on  a cylinder; there,
frustration  arises  because of  the  difficulty  of satisfying  local
packing requirements \cite{JJ,perez}. Here, instead, we find many high
density, strongly  bound and nearly degenerate  low energy structures,
which do satisfy local bonding requirements.

We have obtained the (ground state) energy results from the following procedure. For any assumed values of the ring number and cylinder radius, the energy $E_{\nu}(a,R)$ is evaluated by summing contributions from all interatomic interactions:

\be
E_{\nu}(a,R) = \sum_{i<j} U(r_{ij})
\ee

 Here, the sum includes all pairs of atoms, separated by a 3D 
distance $r_{ij}$. It is convenient to measure distances in units of the hard core parameter, to permit scaling between different solutions. Thus, R* and $a^*=a/\sigma$ are reduced distance variables; similarly 
$E^*=E/\epsilon$ is a reduced energy. As an example ($\nu=4$) revealing the energy's 
dependence on these lengths, a contour plot of the function $E_{4}(a^*,R^*)$  is shown in Figure 3. Note that this function possesses two local minima; these correspond to two quite distinct geometries. The minimum with the larger value of $a^*$ (smaller value of $R^*$) corresponds to neighbors within the same ring that are separated by $\Delta r \approx r_{min}$ (the  equilibrium distance of the pair potential). The other minimum energy configuration (larger value of $R^*$) involves nearest neighbors in adjacent rings, separated by $\Delta r \approx r_{min}$ \cite{foot}. While, in either case, the low energy of the structure comes primarily from such optimization of nearest neighbor distances, longer range interactions do play a significant role in determining the total energy. This is evident from the fact that the (reduced) cohesive energy per particle has a maximum value as high as 3.62 for $\nu=4$. This 
is 45 $\%$ greater than the value (5/2 for $\nu=4$) that would result if only nearest neighbor interactions were included.

For each pair of values of $\nu$ and R, one thus determines a unique value $a_{min}^*$ for which this energy function is a global minimum. This optimized value of $a_{min}^*$ appears in the lower panel of Figure 4 and the corresponding energy appears in the upper panel. In the latter, one observes two alternative behaviors: either a single value of R yields a minimum in this function (for $\nu$=1 or 2) or two values yield local minima (for $\nu>2$). In the latter case, with the single exception of $\nu$=3, the lower energy structure is the one with the smaller value of R, i.e., the case with nearest neighbors in the same ring. 

By such an analysis, we derive the ground state energy, shown in Figure 5, representing the global lower bound of the ensemble of curves in Figure 4. This scallop shell-like function manifests the following plausible behavior. For very small R*, the lowest solution corresponds to the case $\nu$=1.
Near R*=0.5, the lowest energy shifts to the $\nu$=2 structure; this is a plausible result because then two atoms may occupy the same ring without significant hard core repulsion. For increasing R, the minimum energy and structure undergo a sequence of transitions between different values of $\nu$. Interestingly, the sequence is not monotonic: after a very narrow region ($0.60 <R^*<0.63$) in which $\nu$=3 is optimal, there occurs a region ($0.63<R^*<0.73$) in which $\nu$=2 is optimal, followed by an extended region ($0.73<R^*<0.87$) in which $\nu$=4 provides the lowest energy. Note that there are many energy minima close to the (reduced) energy $E/(N\epsilon)=-3.6$. The global minimum energy structure occurs at R*=0.78, with reduced energy -3.62. At that point, there are two neighbors at reduced distance 1.10 and four at distance 1.11, both of which are very close to the pair potential minimum value, 
$r_{min}^* \approx 1.12$. This most cohesive configuration corresponds to a cohesive energy some 7 $\%$ higher than the two-dimensional ground state energy (3.38) of the L-J potential (corresponding to a hexagonal packing) \cite{book}. This result implies that if atoms were to self-assemble on a surface of any shape, the cylindrical surface would be stable relative to the planar surface. While we have not performed the corresponding calculation for atoms on a spherical surface, we suspect that the energy in that case would be competitive with the present results \cite{bal}. Based on experience found in the case of a Coulomb interaction \cite{JJ,perez}, we expect that frustration due to many energetically similar structures would be likely to occur in the spherical case. We note, for completeness, that the reduced cohesive energy in 1D is slightly greater than one (1.03), while in 3D the value is quite large (8.1), a result of both the higher coordination possible in 3D and the large contribution of long range forces.

A structure of the type we are studying will sustain sound waves with various polarizations. The simplest such wave is a longitudinal compressional wave, with propagation vector parallel to the cylinder's axis. In the long wavelength limit, the corresponding sound speed for the case of mass M particles will satisfy:

\be
M s^2 = a^2 \frac{d^2(E/N)}{da^2}
\ee

Here the derivative is evaluated at the ground state configuration for any R. Figure 6 shows corresponding energy curves from which the derivatives in Equation 2 may be computed. For the curves shown, the reduced sound speed $Ms^2/\epsilon$ has the values 86, 37 and 91 at R*=0.78, 1.05 and 1.25 (the three lowest energy minima in Figure 6), respectively. In the case of a mass 16 particle (CH$_4$), these values correspond to $s \approx 3$ km/s, comparable to the bulk speed of sound of CH$_4$. The very high value is indicative of a very tightly bound and rigid structure.

\section{Thermodynamics} 

There exists a venerable tradition of applying the (Ising) lattice gas model to describe the condensation of gases. The critical exponents of the liquid-vapor transition are believed to be exactly determined with this model. As is common in such applications, we simplify our calculations by assuming that only nearest neighbors interact and these interactions all have the same energy ($-J$). While $J$ might simply be set equal to the well-depth of the pair potential, a more sophisticated model might increase the value of $J$ to incorporate attractive, longer range interactions in some approximation. The previous section's results for the energy would suggest that an increase of $J/\epsilon$ by a factor $\sim$ 1.5 is needed to derive the ground state's energy. However, at a density less than complete filling of sites, the occupation fraction would reduce this hypothetical long range correction significantly. Of course, accurate calculations (not undertaken here) would incorporate longer range interactions explicitly in the Hamiltonian itself.

Our method of study is the explicit evaluation of the partition function, within the canonical ensemble. For the case when N sites are occupied at temperature T, this function is
\be
Q_N(T)=\sum e^{-E\{n_i\}/T}
\ee
Here, Boltzmann's constant is taken to be one, the sum is over all configurations $\{n_i\}$ that yield a total of N particles (out of $N_s$ sites), and $E\{n_i\}$ is the corresponding energy. Each configuration corresponds to a specific choice of occupied sites. Periodic boundary conditions are employed, so that the right end sites of a periodic region interact with ``neighboring'' left end sites of that region. A typical calculation involves a configuration determined by the occupancy of the 6$\nu$ sites contained within three unit cells of the lattice. We explore the accuracy of this procedure by varying the size of the periodic cell. Because there is no transition in this 1D system, the finite size effects do not attenuate any divergence in the specific heat, but they are observable. Checks on the results come from the known ground state energy and the high T energy, obtained from a random site occupancy:
\be
E/(NJ) = -\frac{\gamma}{2} \frac{N}{N_s}=-\frac{\gamma}{2} \theta
\ee
Here $\gamma$ is the coordination number and $\theta$ is the occupied 
fraction of sites. Another check comes from the entropy $S(T)$, which is obtained by integrating the heat capacity, divided by T, from zero to infinity, where $S=\ln\{N_s!/[N! (N_s-N)!]\}$. Note that $S/N_s= - \theta \ln \theta - (1-\theta) \ln(1-\theta)$, i.e. ln 2 at 50 $\%$ occupancy.

Figures 7 to 9 show the energy and specific heat as functions of the reduced temperature T*=T/J for the cases $\nu$=2, 3 and 4, respectively. 
We make a number of remarks about the results. First, the curves are all qualitatively similar; this is not surprising because none of these finite $\nu$ cases exhibits a phase transition. Hence, all of the interesting behavior is concentrated in the regime near and below T*=1. The key qualitative dependence is a concentration, with increasing $\nu$ of the thermal ``activity'' into a progressively narrower region of T. This trend is plausible because the limit of very large $\nu$ is the (triangular lattice) 2D limit, which exhibits a genuine phase transition, with critical temperature $T_c^* \approx 0.91$.

Size effects are present in our calculations and may also occur in nature, where nanotubes are finite or may have finite segments that are perfectly ordered. To explore these effects, we focus on the case $\nu=3$. Figure 10 shows how the ground state energy (inset in left panel), at $\theta$=1/2, depends on $N_s$, the number of sites in the unit cell, while the high T result remains fixed at $E/(NJ)=-3/2$, the exact result at high T arising from the 50 $\%$ site occupancy. The dependence on $N_s$ for $\theta$=1/2 conforms to the expression $E(T=0)/(-NJ)= 3 - 12/N_s$, derived from the bulk and surface energies of a ground state ``island'' consisting of $N_s/6$ isolated rings. Hence, the largest system shown ($N_s=30$) has a ground state energy differing from the infinite system result by a fraction 4/30 $\approx$ 0.14. The specific heat bump moves to progressively higher T as $N_s$ increases; it is seen to be converging to a well defined limit at that point, with a maximum near T*=0.89, as is expected from the limiting behavior described above. 

One particularly interesting feature is the presence of a double maximum in the specific heat at fractional occupations $\theta$ both near, but not equal to, 1/2 (Fig. 10, right panel). Note that this behavior becomes increasingly evident as the system size grows, indicating that it is not a finite size artifact. To explore this phenomenon, we compute two correlation functions, defined in the following way: The transverse correlation function is obtained from the average of the product of the occupation numbers of the three sites in the same ring, and the longitudinal correlation function is the average of the product  of the occupation numbers of three sites in consecutive rings.

These correlation functions are plotted in Figure 11 in the case of $\nu$=3, for the same occupations as in Fig. 8. One observes drastically different behavior for $\theta = 1/2$ and  $\theta \neq 1/2$. For $\theta = 1/2$, there arises transverse correlation below T*=0.75, persisting down to T=0, while the longitudinal correlations remains low and constant for the whole range of temperature. The ground state includes completely filled rings (perfect transverse order) but the longitudinal order is imperfect, e.g. because the islands have edges at $\theta = 1/2$. Note that the peak in the specific heat appears at the same temperature (T*$\approx$ 0.75) for which the transverse correlations start to develop, indicating a quasi transition to a more ordered state along the azimuthal direction.

How do we explain the second, low T, bump in C for $\theta$ close to 1/2 ? We attribute it to development of the longitudinal correlation at a lower T since the rings are not all complete in this case. The upper panels of Figure 11 ($\theta = 8/18$ and 10/18) show how the longitudinal correlation arises for T* below 0.25, i.e. a much lower T than where the transverse order begins, giving rise to the corresponding low T bump in Figure 8. Rings start to form at T* $\approx$ 0.75 and this ordering leads to a longitudinal quasi-condensation when these rings start to order along the z direction as T is lowered. When the occupation is considerably larger that 1/2 (for example as shown in the bottom right panel of Fig. 11, for $\theta = 15/18$), the longitudinal order prevails because almost all the rings are occupied and the low T islands are bigger.

Based on the behavior of these two correlation functions we can also explain the absence of two peaks in the $\nu=2$ and $\nu=4$ cases. For half filling, the longitudinal order is always poor because the system tends to aggregate in an island and the peak in the specific heat corresponds to the development of the azimuthal order characteristic of this structure. For occupancy much greater than 1/2, the system tries to condense along the z direction, which coincides with the formation of a much bigger island and the specific heat peak indicates the the appearence of this longitudinal order. However, because the energy difference between the transverse and longitudinally ordered structures is smaller than in the $\nu=3$ case, for intermediate cases when $\theta$ is close to 1/2, longitudinal and transverse ordering occur at nearly the same T, resulting in a single peak in the specific heat.

\section{Quantum effects}

The preceding discussion deals with the classical lattice problem,
leaving open the question of quantum effects. If one were to compute
the rms fluctuations of atomic displacements due to zero-point motion,
the (essentially) 1D phonon states of this lattice would lead to a 
divergent result.
Hence, no crystalline state is possible for our lattice, even at T=0.
However, as in the analogous 2D problem at finite T, the divergence
might not eliminate the possibility of a solid phase, exhibiting
topological long range order (as distinct from  crystalline order, i.e.
periodicity)\cite{kost,nelson}. We do not speculate further about 
this problem here.

However, we do wish to consider the implication of our results for the
case of a rather extreme quantum problem, the ground state of $^4$He atoms
confined to a cylindrical surface. In this case, there is no ambiguity about the ground state, since it is surely a liquid, as in 3D. Our 
concern here is a quantitative
one: what is the binding energy of this ground state? We
answer this question with an estimation method similar to one used by
F. London more than half a century ago in describing $^4$He in 3D \cite{lon}. The goal here is to see whether the lattice model contains any physics relevant to the $^4$He case.

One estimate of the total energy $E_{tot}$ of the system is the following:
\be
E_{est}^{(1)} = <V>_0 + K_{est} 
\ee

Here $<V>_0$ is the ground state energy computed in Section II for a
cylindrical lattice. The latter is a lower bound to the true
potential energy of the quantum problem. Hence, the total $E_{est}^{(1)}$ is a lower bound if the 
kinetic
energy $K_{est}$ were known to be less than, or equal to, the true 
kinetic energy of $^4$He.
Unfortunately, we cannot establish such a relation without carrying 
out a full calculation. Instead, we approximate $K_{est}$, following
London, leading to a total energy estimate, instead of a genuine bound. 
Our estimate of the kinetic energy per particle includes contributions 
from both azimuthal and longitudinal motions for atoms within the lattice:
\be
K_{est}/N = \frac{\hbar^2}{2m}[(\frac{\pi}{a})^2 +(\frac{\nu}{R})^2]
\ee

Figure 12 shows the results of this calculation, along with several others. One of these is a variational calculation of the ground state energy $E_{var}$ of liquid $^4$He on a cylindrical surface, by Carraro \cite{var}. While this latter is just an upper bound to the exact energy, it probably comes within 0.2 K of the exact energy. In comparing these curves, we note that both the energy estimates $E_{est}$ and $E_{var}$ exhibit a common feature: there is a single total energy minimum in the vicinity of 
R* $\approx$ 0.7 to 0.8. We observe that $E_{est}^{(1)}$ lies significantly below Carraro's results for all values of R*. The large discrepancy ($\approx$ 
15 K) is attributable, in our opinion, primarily to the neglect of the large increase in potential energy above the classical lattice energy $<V>_0$. To analyze this, we make a drastic approximation: that the motion of the system is totally harmonic, i.e. a phonon description. In that case, the potential energy increase above the value $<V>_0$ is equal to the kinetic energy per particle. This leads to a revised estimate of the energy:
\be
E_{est}^{(2)} = <V>_0 + 2 K_{est}
\ee

As seen in Figure 12, the variational results lie midway between $E_{est}^{(1)}$ and $E_{est}^{(2)}$ and the minima of these functions lie close to that of the variational calculations. This comparison suggests that our model is getting the physics approximately correct. However, one must bear in mind that a more complete analysis would take into account the relaxation of the system to incorporate the kinetic energy. This would be the analog of allowing the liquid density to vary in searching for the ground state, as is conventionally done in variational calculations (and was done by London with his analogous model). The result of such an approach would be a reduction in the density (as occurs for 3D $^4$He and H$_2$) and in the magnitudes of all energies in the system. Such an extended analysis seems unwarranted in view of the naivet\'{e} of the present description and the availability of alternative, more accurate, computational methods.

\begin{acknowledgments}
We are grateful to Carlo Carraro for helpful discussion. This research has been supported by the Petroleum Research Fund of the American Chemical Society and by Fundaci\'on Antorchas.
\end{acknowledgments}

\newpage

\draft

\begin{figure*}
\includegraphics[height=5in]{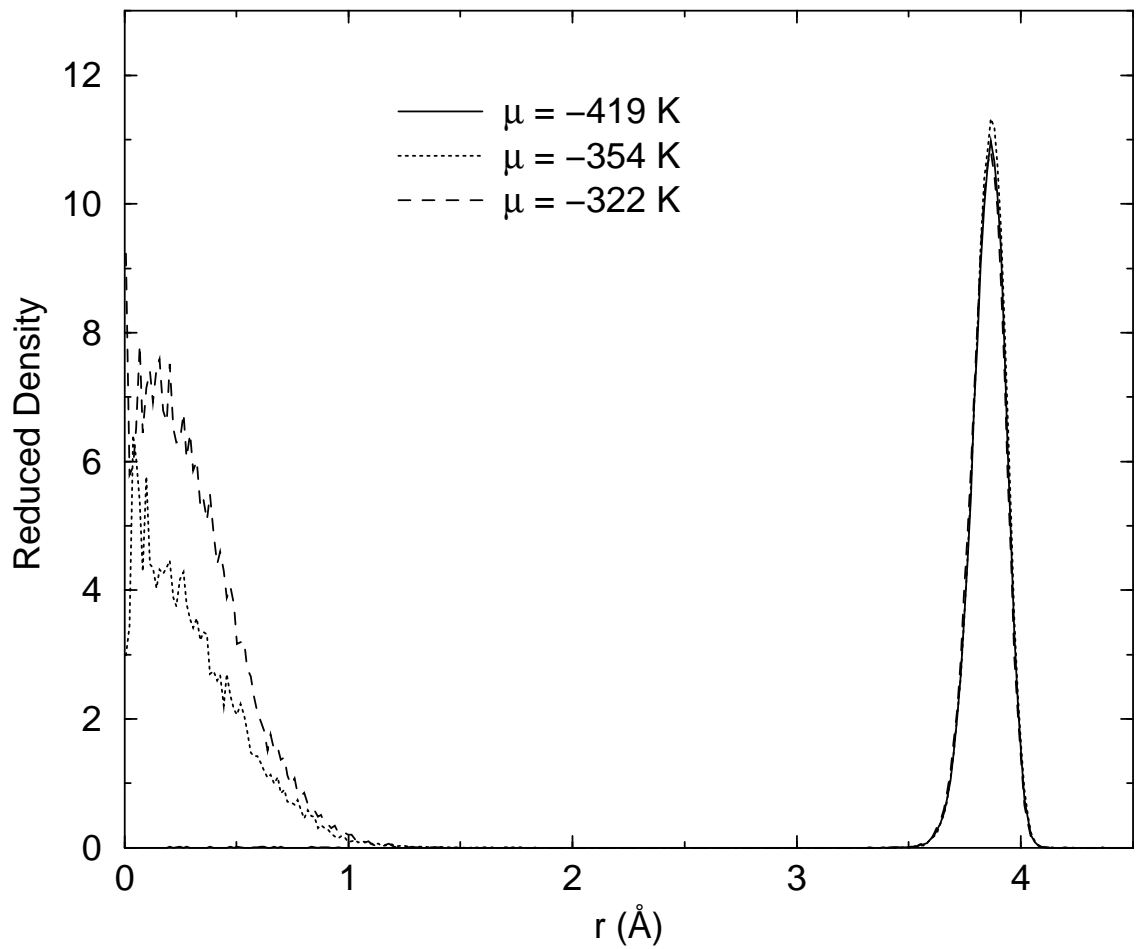}
\caption{Density of H$_2$ molecules at T=10 K as a function of radial distance inside a nanotube of radius 7 $\,$\AA$\,
$ for several values of the chemical potential $\mu$ (taken from Ref. [2]).}
\end{figure*}

\begin{figure*}
\includegraphics[height=7in]{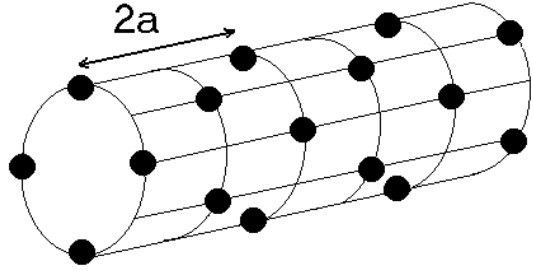}
\caption{Schematic depiction of the cylindrical lattice structure having ring number $\nu$=4. 
The lattice constant is 2a.}
\end{figure*}

\begin{figure*}
\includegraphics[height=5in]{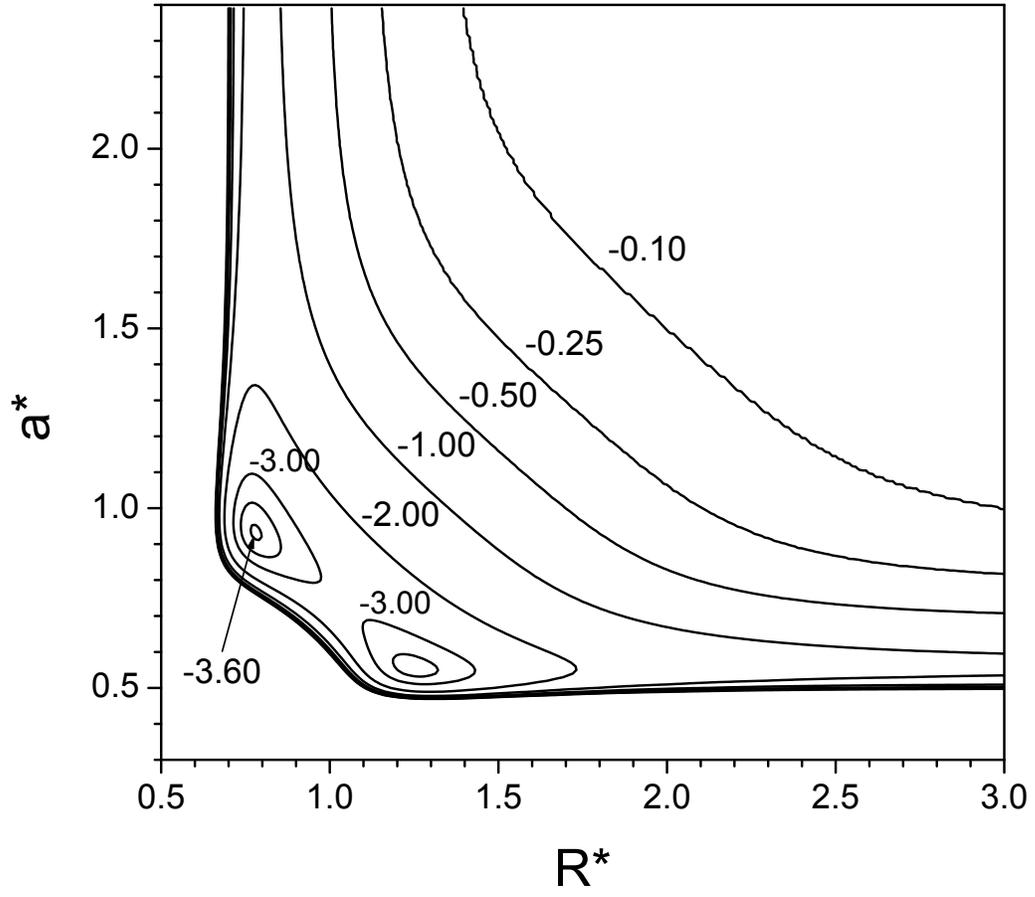}
\caption{The energy per particle $E_{\nu}(a^*,R^*)/(N\epsilon)$ is shown for the case $\nu$=4, as a function of the ring separation a* and radius R*.}
\end{figure*}

\begin{figure*}
\includegraphics[height=5in]{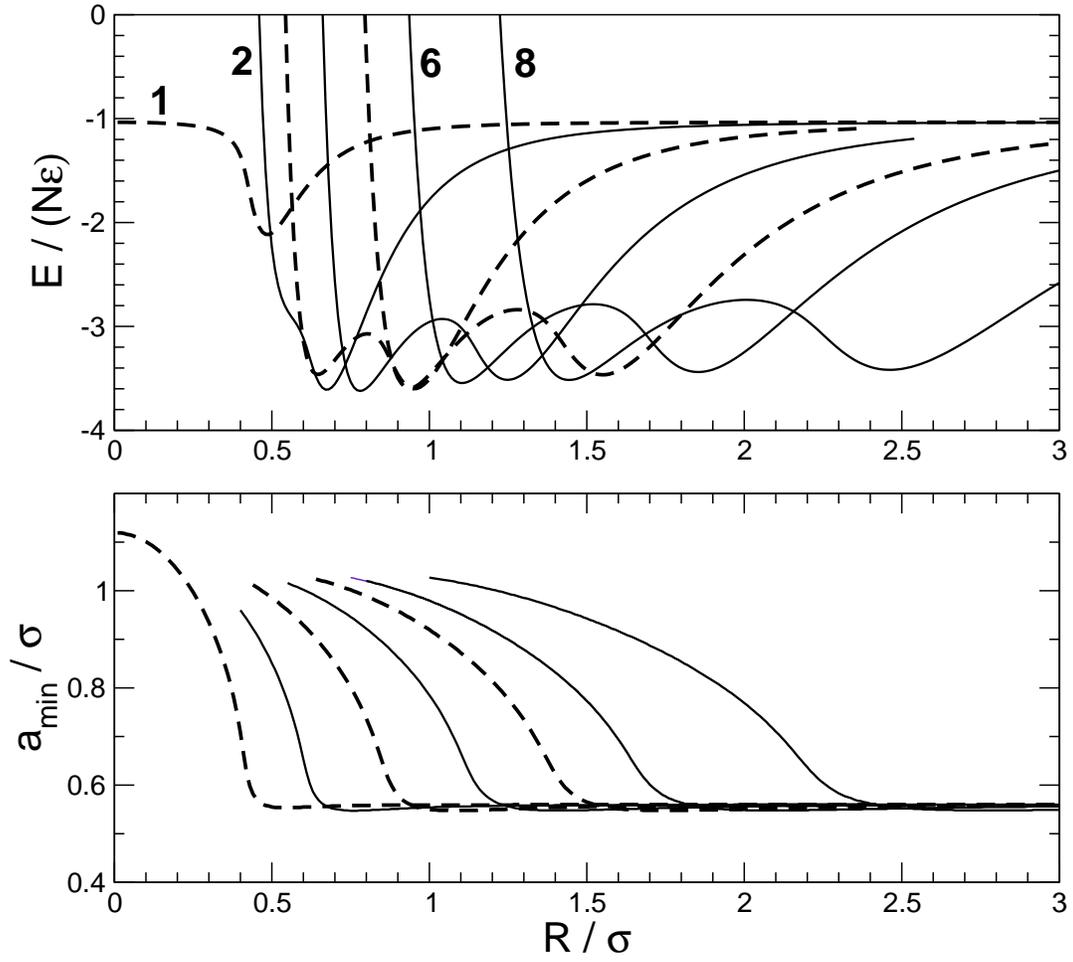}
\caption{Upper panel shows the energy per particle (in units of the pair potential's well depth) as a function of reduced radius, for various assumed ring numbers, $\nu$=1, 2, 3, 4, 5, 6, and 8, from left to right. Each value of E is derived by choosing the optimized lattice constant ($a_{min}$), shown in the lower panel. The odd $\nu$ curves are dashed and $\nu$=1, 2, 6 and 8 are labelled.}
\end{figure*}

\begin{figure*}
\includegraphics[height=5in]{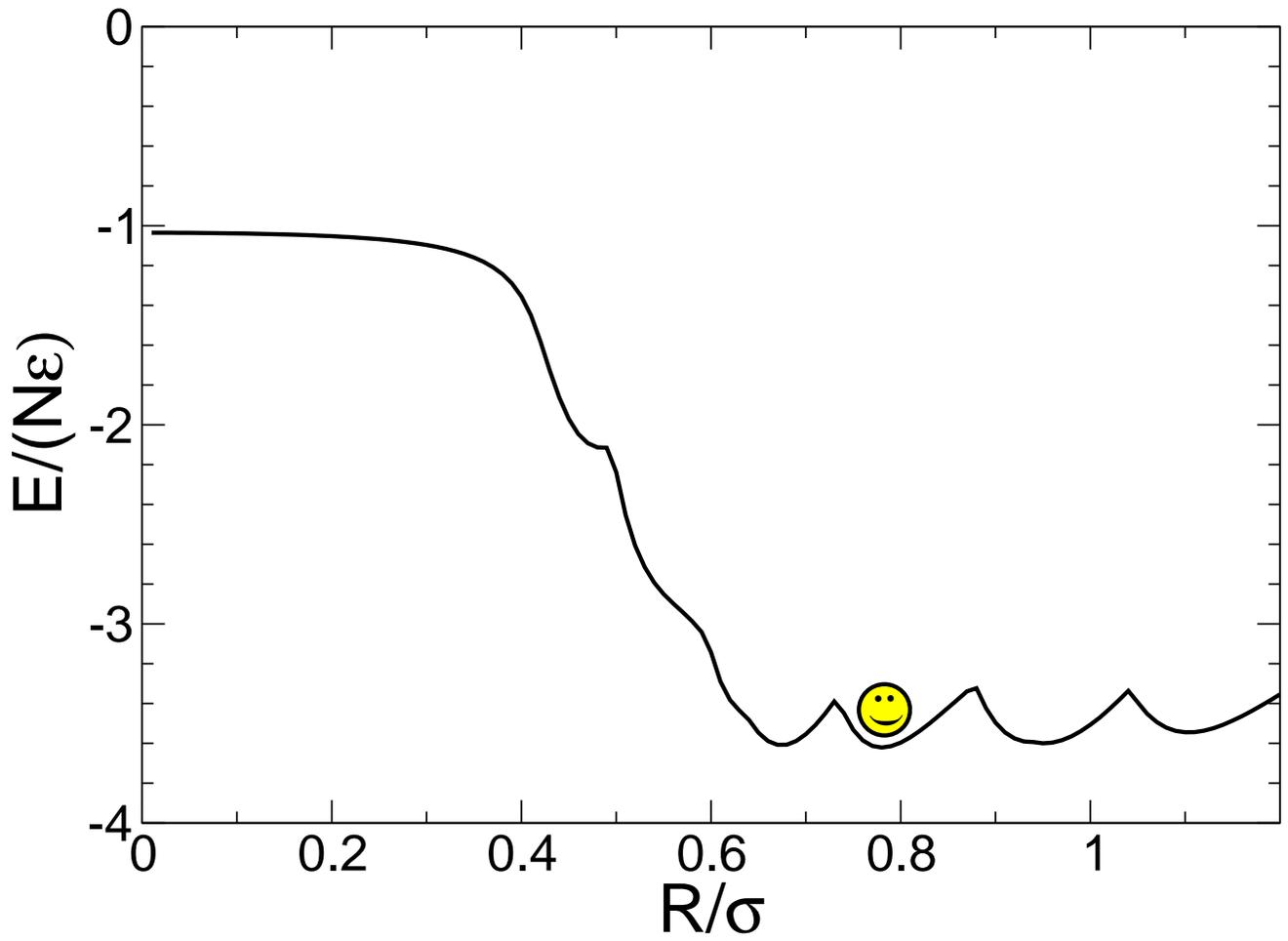}
\caption{Energy as a function of radius, obtained by selecting the minimum energy from the solutions in Figure 4. The global minimum energy for this problem occurs at R*=0.78 ($\nu = 4$), as indicated by the happy face.}
\end{figure*}

\begin{figure*}
\includegraphics[height=5in]{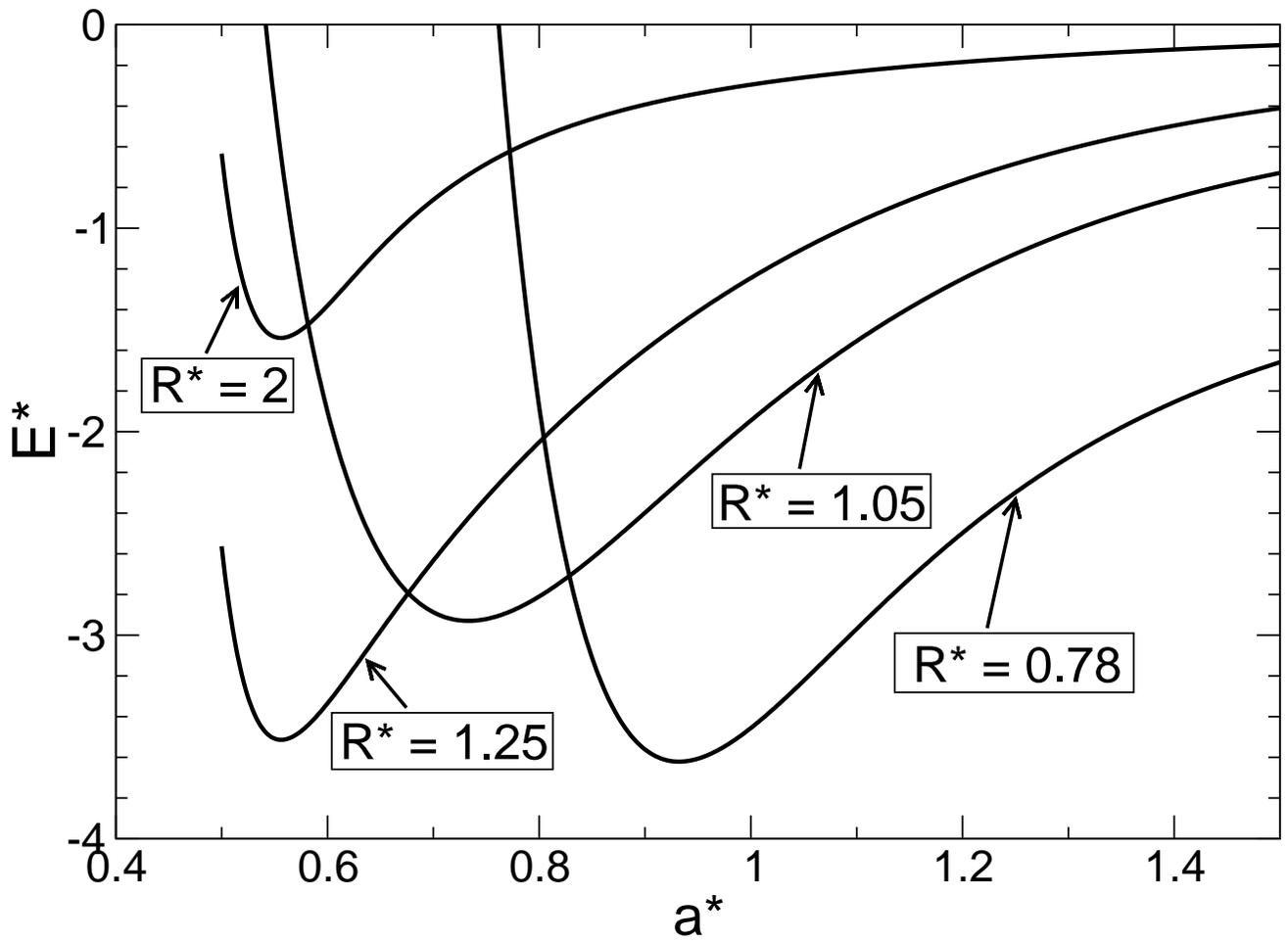}
\caption{Energy as a function of a* for various values of R* in the case $\nu=4$, see Fig. 3. The curve corresponding to R*=0.78 yields the global energy minimum of the problem.}
\end{figure*}

\begin{figure*}
\includegraphics[height=5in]{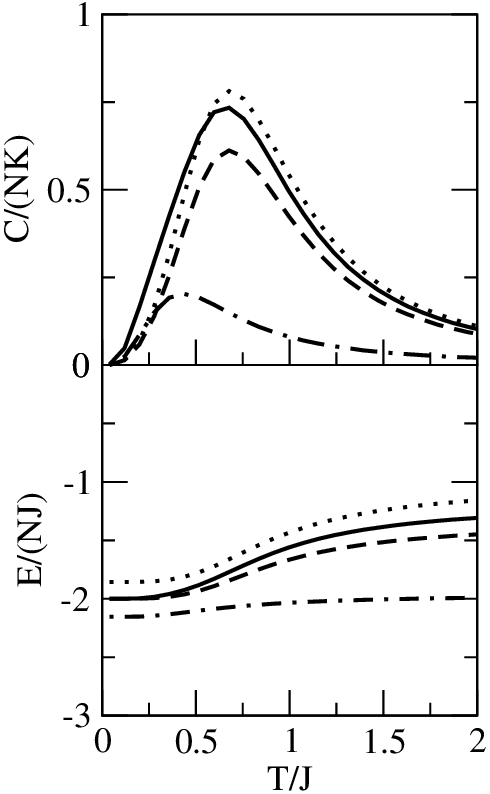}
\caption{Energy (lower panel) and specific heat (upper panel) for the case of ring number $\nu=2$, $N_s = 16$ (8 rings). The full curve corresponds to half filling occupancy ($\theta$=8/16), the dotted curve to $\theta$=7/16, the dashed one to $\theta$=9/16, and the dot-dashed to $\theta$=13/16.}
\end{figure*}

\begin{figure*}
\includegraphics[height=5in]{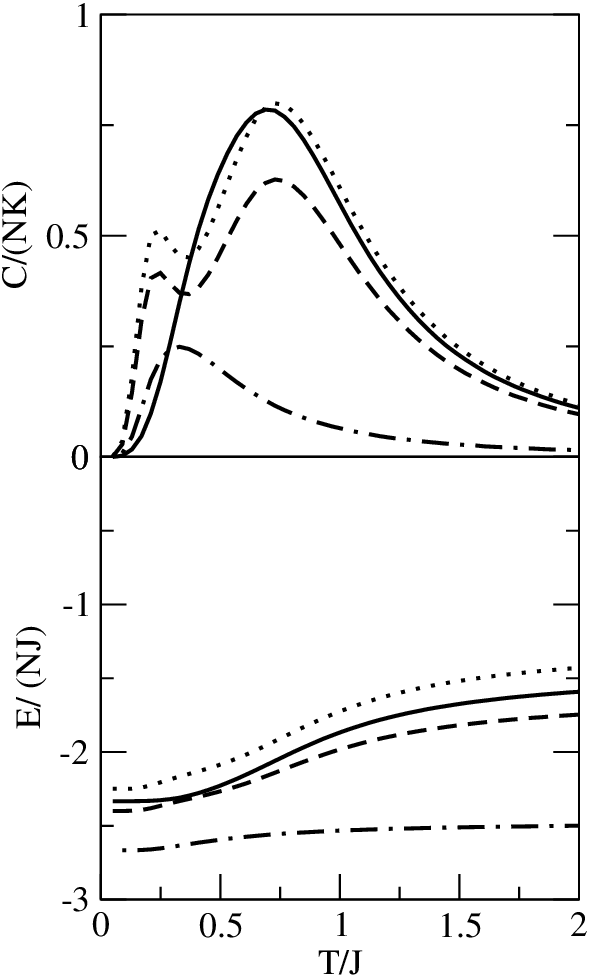}
\caption{Energy (lower panel) and specific heat (upper panel)for the case of ring number $\nu=3$, 
$N_s = 18$ (6 rings). The full curve corresponds to half filling occupancy ($\theta$=9/18), the dotted curve to $\theta$=8/18, the dashed one to $\theta$=10/18, and the dot-dashed to $\theta$=15/18.}
\end{figure*}

\begin{figure*}
\includegraphics[height=6in]{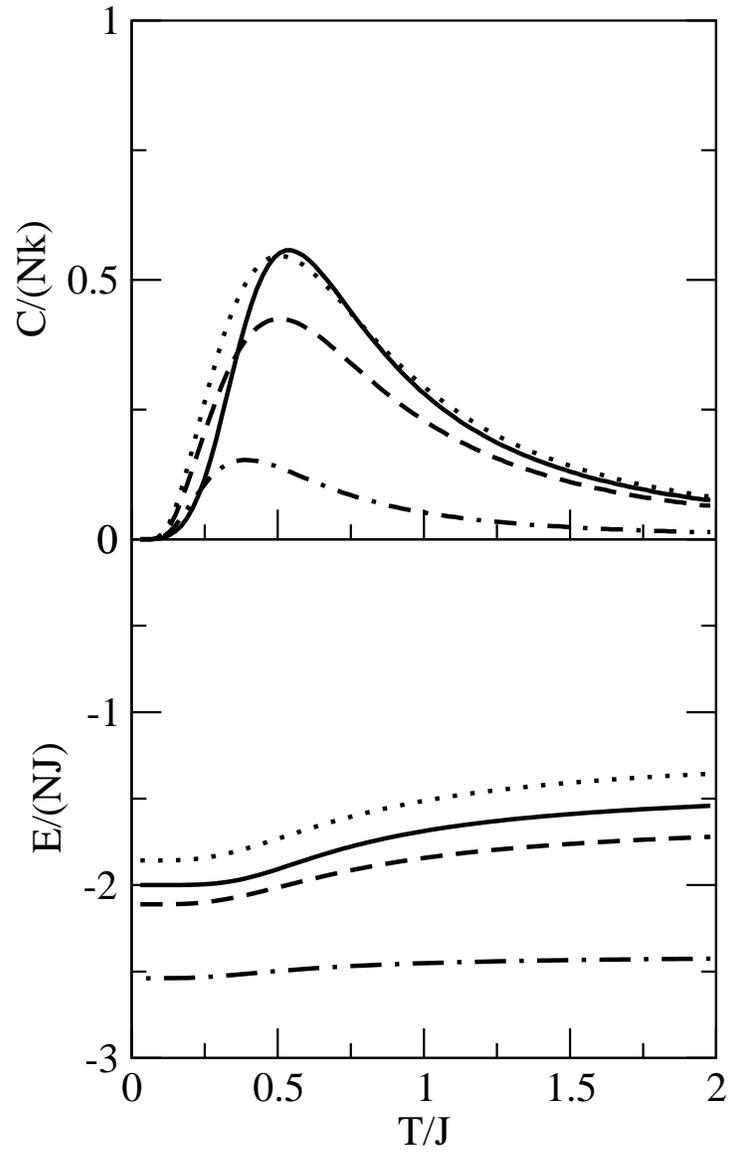}
\caption{Energy (lower panel) and specific heat for the case of ring number $\nu=4$, $N_s = 16$ (4 rings). The full curve corresponds to half filling occupancy ($\theta$=8/16), the dotted curve to $\theta$=7/16, the dashed one to $\theta$=9/16, and the dot-dashed to $\theta$=13/16.}
\end{figure*}

\newpage

\begin{figure*}
\includegraphics[height=3in]{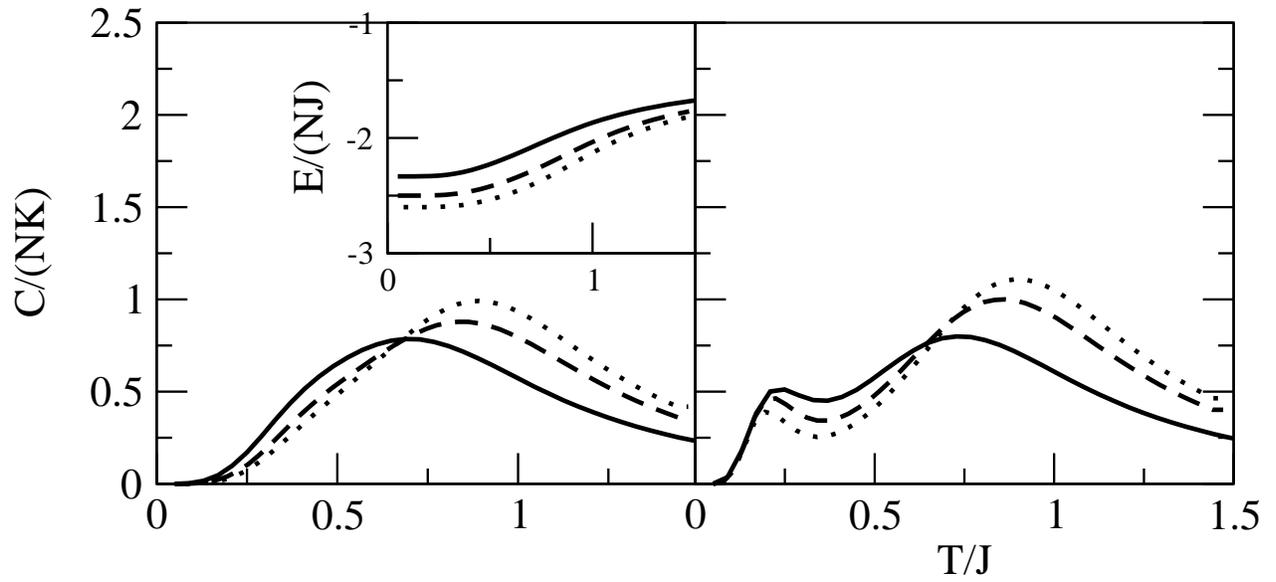}
\caption{Specific heat for $\nu=3$, showing the size effect as the number of sites $N_s$ increases. The full curve depicts the case for $N_s=18$ (6 rings), the dashed curve corresponds to $N_s=24$ (8 rings) and the dotted curve to $N_s=30$ (10 rings). The left panel show the case for half occupancy (the inset depicts the energy per particle) whereas the right panel corresponds to occupancy below 1/2: $\theta =$ 0.44 (full curve), 0.42 (dashed curve), 0.43 (dotted curve).}
\end{figure*}

\begin{figure*}
\includegraphics[height=5in]{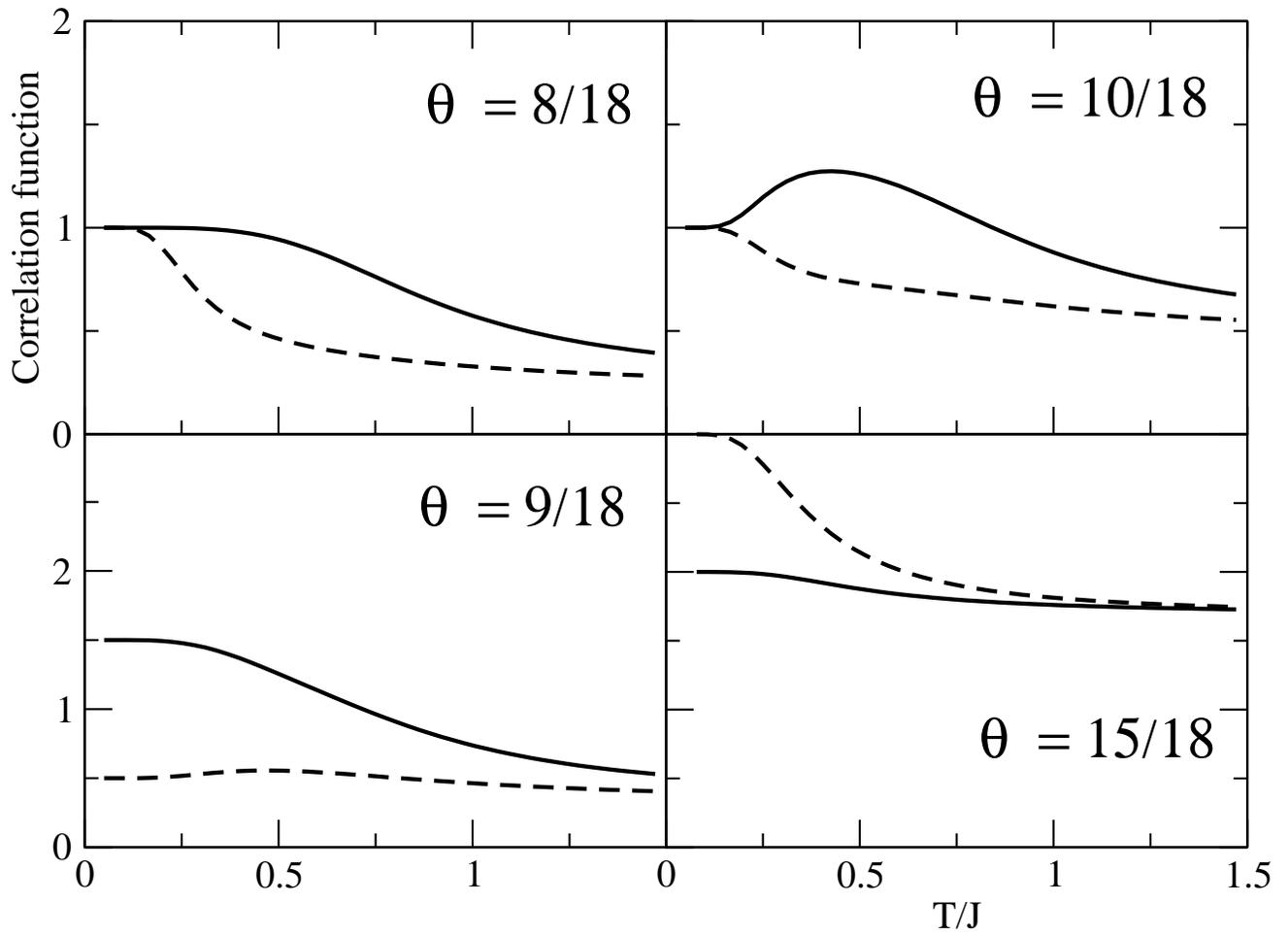}
\caption{Transverse (full curves) and longitudinal (dashed curves) correlation functions (defined in the text) for the same occupancies as in Fig. 8 ($\nu=3$).}
\end{figure*}

\begin{figure*}
\includegraphics[height=5in]{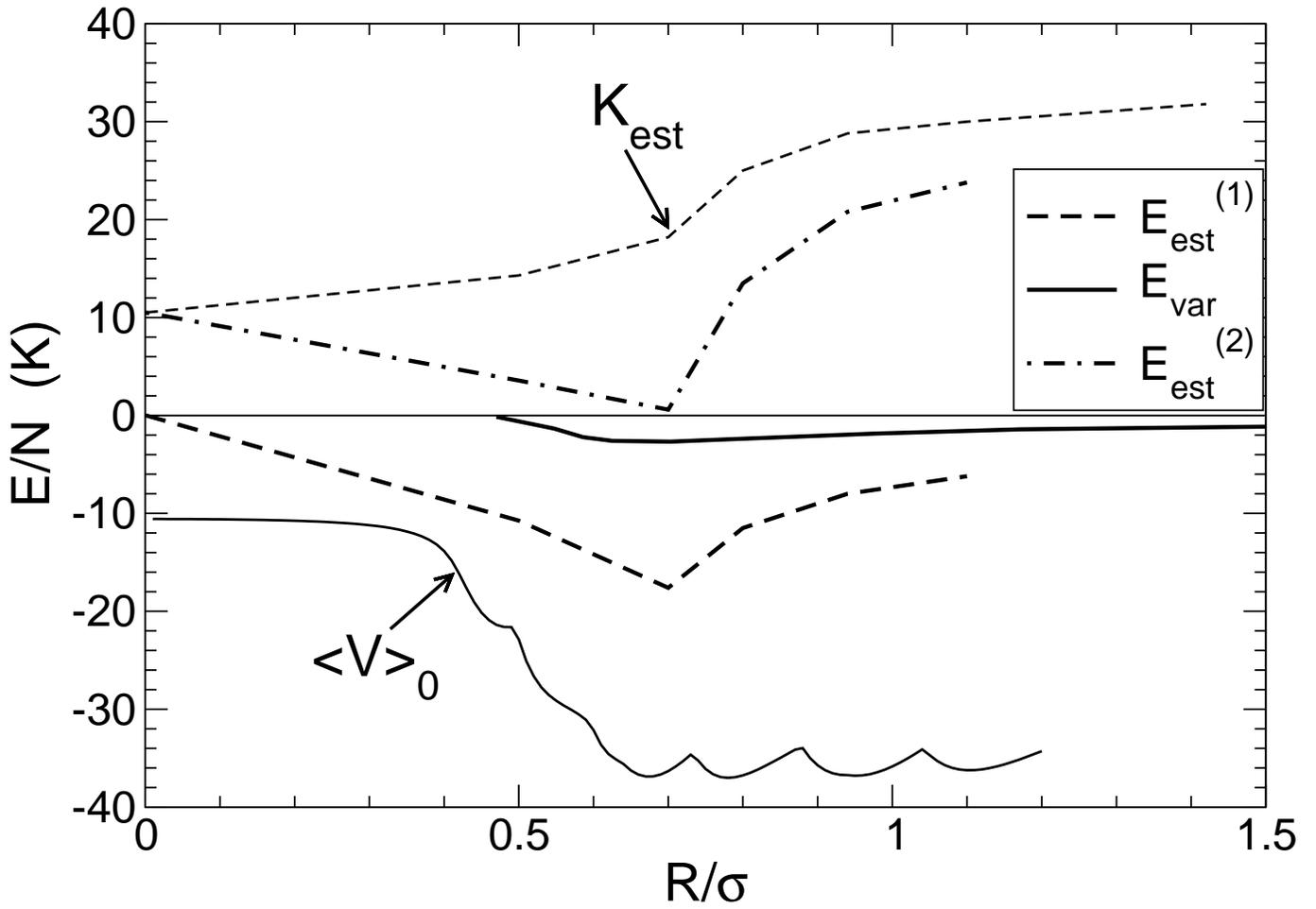}
\caption{The ground state energy of $^4$He atoms confined to a cylindrical surface.
The full curve shows the result of the variational calculation $E_{var}$ by 
Carraro (Ref. 1). Also shown are two alternative estimates $E_{est}^{(1)}$ and $E_{est}^{(2)}$ of the total energy obtained from the classical potential energy $\rangle V \langle_0$ and estimated kinetic energy, K$_{est}$, as described in the text.}
\end{figure*}


\begin{references}

\bibitem{rivista} Mary J. Bojan, M. Mercedes Calbi, Carlo Carraro, 
Milton W. Cole,
Silvina M. Gatica, M. L. Glasser, E. Susana Hernandez and
Milen K. Kostov, {\em Matter on a Cylindrical Surface}, to be submitted 
to Rivista del Nuovo Cimento.

\bibitem{axial} S. M. Gatica, G. Stan, M. M. Calbi, J. K. Johnson, and
M. W. Cole, J. Low Temp. Phys. {\bf 120}, 337-360 (2000).

\bibitem{green} D. Green and C.
Chamon, Phys. Rev. Lett. {\bf 85}, 4128 (2000).

\bibitem{millot} F. Millot, Y. Larher, and C. Tessier, J. of Chem. Phys.
, {\bf 76}, 3327, 1982; V.L. Pokrovsky and A.L. Talapov,{\em Theory of
incommensurate crystals}, Harwood Academic Publishers, London (1984).


\bibitem{milen} M.K. Kostov, M.W. Cole, J.C. Lewis, P. Diep and J.K. Johnson, Chem. Phys. Lett. 332, 26-34 (2000).

\bibitem{JJ} J.J.Thomson, Phil. Mag.7, 237 (1904). 

\bibitem{perez} A. Perez-Garrido and M.A. Moore, Phys. Rev. B {\bf 60}, 
15628 (1999); M. J. Bowick, D. R. Nelson and A. Travesset, Phys. Rev. B 
{\bf 62}, 8738 (2000).

\bibitem{foot} To be specific, in the case $\nu$=4, the nearest neighbor separations r* for the smaller R* minimum are 1.11 for atoms in the same ring and 1.10 for atoms in adjacent rings (compared to the equilibrium 
value 2$^{1/6} \approx$ 1.12). For the larger R* case, the atoms in the same ring are separated by 1.77 while those in adjacent rings are separated by 1.11.

\bibitem{book} L.W. Bruch, M.W. Cole, and E. Zaremba, {\em Physical Adsorption: Forces and Phenomena}, Oxford University Press (1997).    

\bibitem{bal} M. K. Balasubramanya and M. W. Roth, Phys. Rev. B {\bf 63}, 205425 (2001); M. W. Roth and M. K. Balasubramanya, Phys. Rev. B {\bf 62}, 17043 (2000). 

\bibitem{kost} J.M. Kosterlitz and D.J. Thouless, {\em Progress in Low 
Temperature Physics}, Vol. 7B, pp.371--433, North Holland, New York (1978).

\bibitem{nelson} D.R. Nelson, in {\em Phase transitions and critical phenomena}, Vol. 7, ed. C. Domb and J. L. Lebowitz, pp.1--99, Academic Press, 
New York (1983).

\bibitem{lon} F. London, {\em Superfluids}, Vol. II, Dover Publications,
New York (1954).

\bibitem{var} C. Carraro, unpublished.

\end{references}
\end{document}